# Adsorption Structure Determination of a Large Polyaromatic Trithiolate on Cu(111): Combination of LEED-I(V) and DFT-vdW†

Thomas Sirtl,[a] Jelena Jelic,[b] Jörg Meyer,[b] Kalpataru Das,[c] Wolfgang M. Heckl,[a,d] Wolfgang Moritz,[e] John Rundgren,[f] Michael Schmittel,[c] Karsten Reuter,[b] and Markus Lackinger[a,d,*]

The adsorption geometry of 1,3,5-tris(4-mercaptophenyl)benzene (TMB) on Cu(111) is determined with high precision by two independent methods, experimentally by quantitative low energy electron diffraction (LEED-I(V)) and theoretically by dispersion corrected density functional theory (DFT-vdW). Structural refinement by both methods consistently results in similar adsorption sites and geometries. Thereby a level of confidence is reached that allows deduction of subtle structural details such as molecular deformations or relaxations of copper substrate atoms.

## Introduction

The atomically precise structure determination of large functional organic adsorbates on surfaces is a challenging task in surface science. Abundantly used scanning tunneling microscopy (STM) yields unit cell parameters for molecular superstructures with a typical accuracy of 5%. Normally it is also possible to deduce the azimuthal orientation of larger adsorbates with respect to the surface. However, already the determination of adsorption sites can become intricate.[1] Under favourable circumstances estimation of adsorption heights and molecular deformations may be feasible by STM,[2] a precise quantification remains impossible. Such details are important for a fundamental understanding of the interactions and properties of adsorbed molecules though. For instance deformations can affect the aromaticity of conjugated molecules and also change electronic properties that are decisive for applications.[3]

A corresponding, more quantitative determination of the internal adsorption geometry is to date the realm of diffraction methods. Vertical adsorption distances are accessible by x-ray standing wave (XSW) experiments.[4,5] However, besides the lack of lateral resolution, a further drawback of XSW is that for a specific element in comparable chemical surrounding only averaged height data can be obtained. This restriction does not apply to quantitative low energy electron diffraction, LEED-I(V), which is furthermore an experimentally much less elaborate technique.[6] Here the intensities of unique reflections in a LEED experiment are recorded as a function of electron energy. A prerequisite for this diffraction technique is the availability of long-range ordered monolayers, and owing to the elaborate nature of the scattering simulations application of the technique was hitherto restricted to smaller, conformationally rigid adsorbates like dinitrogen,[7] carbon monoxide,[8] formic acid,[9] cyanide,[10] acetylene,[11] glycine,[12] thiouracil,[13] benzyne,[14] or benzene.[15-20] LEED-I(V) analyses of larger adsorbates to date are rare, examples being studies on graphene[21] and C$_{60}$ fullerenes.[22]

Recent advances in computer power and simulation software greatly alleviate this restriction to smaller adsorbates, and thus offer the prospect of atomically precise surface structure determination also of technologically most relevant larger functional molecules. In the present study we illustrate this with LEED-I(V) calculations that were performed with an update of the LEEDFIT code,[23-25] which was parallelized and allowed interatomic distances as constraints in the least squares optimization. A further improvement was the introduction of dynamic phase shift calculations (LEED-PS), where during the structure refinement the changes in phase shifts due to changes in structural parameters and bond lengths are considered by self-consistent recalculation.

On the theoretical side this development finds its counterpart in the advent of numerically most efficient dispersion-correction approaches to density-functional theory (DFT-vdW).[26-28] At essentially zero additional computational cost, these approaches augment the predictive capability of prevalent semi-local DFT functionals with an account of van der Waals interactions, which are known to play a decisive role in determining the structure and stability of organic molecules on solid surfaces.[26-35]

With the present study we demonstrate how the increased performance of both LEED and DFT simulations provide access to surface structural data of complex molecules at sub-atomic precision by studying 1,3,5-tris(4-mercaptophenyl)-benzene (TMB) monolayers on Cu(111). Thiol-functionalized molecules are promising candidates for linkers in molecular electronics and their interaction with metal surfaces is of great interest.[36] On reactive surfaces at room temperature, monothiols deprotonate into thiolates, and on Cu(111) the sulfur head group binds covalently at threefold hollow sites.[37] To date the exact adsorption site of TMB has not been unambiguously identified despite its obvious relevance for the formation of metal-organic coordination networks,[38] *i.e.* it is not clear how the preference for a specific bonding site of the sulfur head groups is matched with the given geometric arrangement of the three thiolate groups in TMB.

## Experimental section

Sample preparation was carried out under ultra-high vacuum by thermal sublimation of TMB onto Cu(111) held at room temperature. Synthesis details of TMB were published elsewhere.[39] LEED experiments were conducted at a sample temperature of 50 K (cf. ESI†). The LEED-I(V) analysis includes 22 unique reflections at normal incidence with electron energies between 11 eV and 200 eV, resulting in a cumulative energy range of 2766 eV. The I(V)-curves were averaged over symmetrically equivalent reflections. The degradation of the reflection intensity during data acquisition due to radiation damage was below 20%. It is noteworthy that the experimentally used electron beam current is a compromise between a sufficiently high signal to noise ratio and low radiation damage.

Phase shifts were derived from a crystal potential obtained by superposition of atomic charge densities. The energy dependent self-energy of the scattered electron was used in the optimization of non-overlapping muffin-tin radii for the atoms of the crystal while minimizing the potential step between the muffin-tin spheres.[40, 41] The same method has been previously applied for oxide surfaces.[42, 43] The optimisation of muffin-tin radii results in a different radius for every combination of atom, crystal, and scattering energy. The phase shifts were therefore iteratively recalculated in the final structure refinement step. The influence of the different methods of phase shift calculation on the structural results will be discussed in a separate paper.

In the final refinement iterations anisotropic atomic displacement parameters were used for the adsorbed molecule.[44] The results show an enhanced rms-amplitude of 0.2 - 0.3 Å compared to 0.05 Å of the substrate atoms, that may be caused by thermal vibration and static displacement due to desorbed hydrogen or other defects. The displacement parameters exhibit large error bars and are not discussed in detail here, because no temperature dependent measurements have been made.

## Results and discussions

Previous STM and LEED experiments of TMB on Cu(111) with similar sample preparation yielded a (3√3×3√3)R30° superstructure with a lattice parameter of 13.3 Å and 27 copper atoms per unit cell in the first layer.[39] From the STM data a fairly large domain size and a low defect density was inferred, rendering the system ideal for surface diffraction studies. The structure exhibits $p31m$ symmetry with one molecule per unit cell and TMB appeared with threefold symmetric submolecular STM contrast. Hence, the likewise threefold symmetric TMB is centred on a threefold symmetric adsorption site, i.e. either fcc or hcp threefold hollow sites or on top. This assumption is also consistent with the observed LEED pattern.

Considering the given $p31m$ symmetry, the asymmetric unit comprises 10 atoms in the adlayer (Fig. 1c). Six distinct adsorption geometries, where the TMB lobes are aligned with a mirror line are consistent with the above stated symmetry requirement. All adsorption geometries are listed in Table 1. Interestingly, the TMB molecule is almost commensurate with the Cu(111) lattice in two respects. In the given azimuthal orientation all four phenyl rings as well as all three peripheral thiolate head groups can simultaneously occupy equivalent adsorption sites without imposing large stress on the molecule.

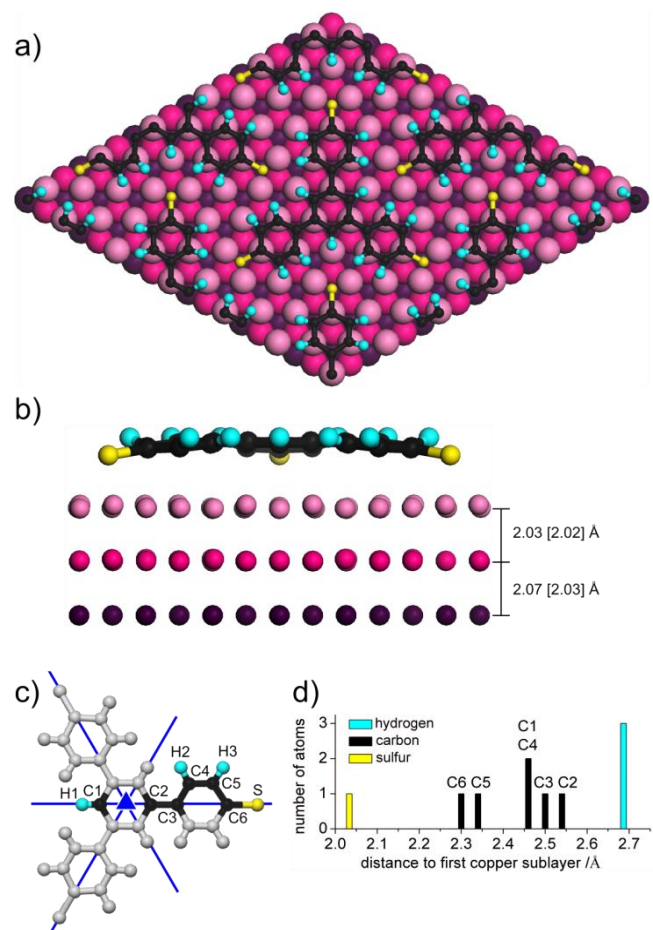

**Fig. 1** Optimized structure of TMB on Cu(111). (a) Top-view of 2×2 unit cells. (b) Side-view of one molecule and three copper layers. Mean distances are depicted for the first three copper layers (the values in parenthesis refer to DFT). (c) Asymmetric unit of TMB (colored). All other atom coordinates (grey) are generated by the symmetry operations of $C_{3v}$ (blue lines, mirror planes; blue triangle, three-fold symmetry axis). (d) Vertical atom distances in the asymmetric unit of TMB, referring to the mean height of the first copper sublayer (LEED-I(V) derived values, for DFT values cf. ESI†, Table S3 and Fig. S3).

**Table 1** LEED-I(V) and DFT results for the six symmetry-allowed adsorption geometries of TMB on Cu(111).

| TMB adsorption site | | LEED-I(V) | DFT | DFT-vdW[a] |
|---|---|---|---|---|
| # sulfur | phenyl | $R_p$ | ΔE /eV | ΔE /eV |
| 1 on top | fcc | 0.74[b] | +1.11 | +1.05 |
| 2 on top | hcp | 0.73[b] | ---[d] | ---[d] |
| 3 fcc | on top | 0.60[b] | ---[d] | ---[d] |
| 4 fcc | hcp | 0.53[b] (0.32)[c] | 0 | 0 |
| 5 hcp | on top | 0.81[b] | ---[d] | ---[d] |
| 6 hcp | fcc | 0.75[b] | +0.55 | +1.40 |

[a] Using a dispersion correction scheme developed by Tkatchenko and Scheffler.[27] [b] Copper atoms fixed. [c] More elaborate refinement of best-fit model. [d] Not optimized until full convergence was achieved.

For both LEED structure refinement and DFT calculations all six adsorption geometries of Table 1 were considered. LEED structure refinement was carried out in the asymmetric unit, i.e.

applying symmetry constraints, while the DFT calculations were conducted without any symmetry constraints. LEED structure refinement was realized in a two-step process. For an initial evaluation of all six structures, constraints for intramolecular distances of TMB were applied (cf. ESI†, Table S1), and three substrate layers were considered. Layer distances, vertical and lateral atom coordinates of TMB were first optimized consecutively, and simultaneously in the final step. Pendry's r-factor ($R_p$) is used to evaluate the agreement of experimental and theoretical I(V) curves.[45] This first step resulted in the unambiguous identification of the actual adsorption site. After identifying the correct model, a more elaborate refinement was performed. To this end, firstly, vertical distances between TMB and three copper sublayers were optimized. Secondly, all atom coordinates of TMB, and subsequently of first and second layer copper atoms were optimized. Since Cu(111) is well known to feature a free adatom gas,[39] with many examples for interference with self-assembly of organic structures,[46-48] two conceivable structures with copper adatoms were considered in addition to those listed in Table 1. Firstly, a structure where each thiolate group binds to clusters of three adatoms was tested, however the best $R_p$ achieved was 0.9 (cf. ESI†, Fig. S5a). Secondly, a structure was tested, where interstitial copper adatoms are adsorbed in the gaps between TMB molecules (cf. ESI†, Fig. S5b). An occupation of 100 % led to an Rp value of 0.6, whereby optimization of the occupation factor together with a full structure refinement of all other parameters led to a local minimum with Rp = 0.42 at an occupancy of 30 % and only marginal modification of the molecule geometry. Both adatom containing structure models were discarded, because the obtained $R_P$ values are significantly larger than that of the best fist model without copper adatoms (vide infra).

Among the six competing structural models (cf. Table 1), structure 4 is unambiguously preferred, in which the three thiolate groups bind to fcc threefold hollow sites and the four phenyl rings reside on hcp threefold hollow sites. Adsorption of phenyl rings on threefold hollow sites is common and was also reported for benzene on Co(0001),[16] Ni(111),[17] and Ru(0001).[19, 20] As evident from the model presented in Fig. 1(a), with the given azimuthal orientation of TMB every other carbon atom resides on top of copper. In the LEED results this preference for structure 4 is primarily expressed by the lowest resulting $R_p$ value of 0.32 (cf. Table 1). Furthermore, only optimization of structure 4 yielded physically reasonable results. Although the alternative structural model 3 exhibited initially a comparable $R_p$ value, the corresponding LEED optimized geometry showed unreasonable distortions of the molecule (cf. ESI†, Fig. S2).

Experimental and calculated I(V)-curves of fully optimized model 4 are in good agreement (Fig. 2), as indicated by an overall $R_p$ of 0.32. We note here that the remaining misfit between the measured and calculated I(V) curves may be partially explained by inadequacy of the muffin-tin approximation used in the multiple scattering formalism. $R_p$ dropped from 0.35 to 0.32 when using the dynamic phase shift adaptation algorithm while the atom positions remained within the error limits of 0.10 - 0.15 Å. Also if the two weakest beams – exhibiting a relatively low signal to noise ratio – are excluded, the $R_p$ value can be further improved to 0.28 without any significant changes in the structure. Nevertheless, the structure discussed on the following is derived from all experimental I(V) curves.

It is commonly agreed that $R_p$ values below 0.2 indicate an excellent agreement between experimental and theoretical I(V) curves, whereas values above 0.3 are interpreted as mediocre fits. Excellent refinements yielding very low $R_p$ values in the range 0.11 - 0.24 were reported for complex inorganic surface structures such as CoO(111),[49] GaN(0001),[50] and $BaFe_2As_2$(001)[51] as well as surface alloys or metal superstructures on metals like Pb/Ni(111),[52] Sn/Ni(110),[53] Sn/Ni(111),[54] Sb/Cu(110),[55] and Au/Pd(100).[56] Moreover, relatively simple atomic superstructures could also be refined with high accuracy, examples comprise hydrogen on Ir(110),[57] ($R_p$ 0.10) and halogens on metal surfaces as Cl/Ru(0001)[58] and Br/Pt(110)[59] with $R_p$ 0.19 and 0.23, respectively. LEED structure refinement was also successfully carried out for sulphide or oxide adlayers like O/V(110),[60] MnO/Ag(100),[61] S/Ir(100),[62] O/Pt/Cu(100),[63] $V_2O_3$/Pd(111),[64] S/Au(110),[65] O/Cu(210),[66] resulting in $R_p$ of 0.11 - 0.36. Excellent $R_p$ values around 0.15 - 0.22 were also obtained for smaller organic adsorbates as dinitrogen/NaCl(100),[7] carbon monoxide/Pt(110),[8] formic acid/$TiO_2$(110),[9] cyanide/Ni(110),[10] acetylene/Cu(111),[11] and glycine/Cu(110).[12]

In this respect, $R_p$ values of 0.27 and 0.29 for thiouracil/Ag(111)[13] and benzyne/Ir(100)[14] seem to indicate a less perfect agreement, but are the state of the art for medium-sized molecules on metal surfaces. Even for a comparatively small and rigid molecule like benzene on Co(0001),[16] Ni(111),[17] Co(10-10),[18] and Ru(0001)[19, 20] relatively large $R_p$ of 0.26 - 0.37 were reported. In the literature $R_p$ values up to 0.40 are thus still considered as reasonable fits for larger adsorbate molecules. In particular for organic superstructures with large unit cells such as graphene[21] or molecules with many atoms such as $C_{60}$ fullerenes on Ag(111)[22] $R_p$ of 0.29 and 0.36 were obtained. Several reasons may contribute to such typically higher $R_p$ values for organic adlayers. On the theoretical side, we demonstrate for the present system that dynamic adaptation of the phase shifts already leads to a significant improvement of $R_p$ from 0.35 to 0.32. Furthermore, for complex structures with large unit cells the commonly made assumptions, i.e. the muffin-tin approximation, isotropic displacement factors, and the neglect of correlations in the displacement factors, might be oversimplifications. In addition, on the experimental side, several reasons might account for lower $R_p$ values. Organic adlayers are much more prone to radiation damage. In addition, low signal to noise ratios of weak reflections also lead to higher $R_p$ values as shown here, where exclusion of the two weakest beams further improves $R_p$ from 0.32 to 0.28. Moreover, structural defects in the adlayer as grain boundaries further infer the quality of the experimental data set. In LEED I(V) analyses in general not the same level of agreement can be reached as it is the standard for example in x-ray powder diffraction. As mentioned before, this results from several approximations used in the multiple scattering theory, namely the phenomenological description of inelastic processes by an optical potential, the neglect of correlations in the atomic displacement parameters and the muffin tin model for single atom scattering. Also defects play a more significant role at surfaces than in bulk samples. Therefore mainly the peak positions are compared in the I(V) curves and an R-factor of Rp = 0.32 is fully acceptable for structure optimization of a large adsorbate.

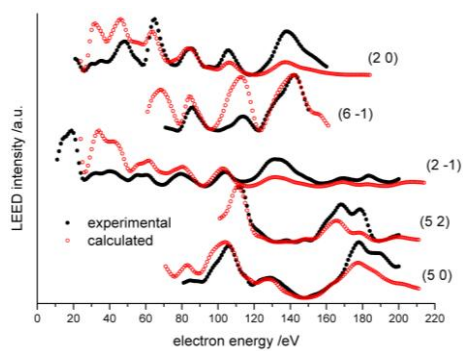

**Fig. 2** Selected experimental and theoretical LEED-I(V) curves for the best-fit model 4 (vertically offset for clarity). Reflection indices are given in brackets. The complete dataset is provided in ESI†, Fig. S1.

The six considered adsorption geometries were also optimized in DFT calculations, using the semi-local Perdew-Burke-Ernzerhof functional.[67, 68] In order to evaluate the significance of van der Waals contributions in a structure that is dominated by covalent anchoring of the thiolate groups, the calculations were conducted with and without dispersion correction.[27, 34] Only three out of the initial six structures were refined until full convergence was achieved. The other three structures were discarded at an earlier stage of the calculation, when the last geometry optimization steps led to energetic improvements on the 10 meV scale and it became clear that their energies will remain significantly higher than those of the three more favourable structures (cf. ESI). The resulting energy differences of these three remaining optimized structures are listed in Table 1. In perfect agreement with the LEED results, structure 4, where the phenyl rings are centred at hcp sites and sulphur binds to fcc sites, yields the lowest energy, both in the calculations with and without dispersion-correction. We take this as an indication for the reliability of the obtained energetic ordering, even though absolute binding energies of prevalent dispersion-corrected DFT approaches are known to be severely impaired at metal surfaces by electronic screening effects.[35]

Intriguingly, the agreement of both independent techniques is not only restricted to the adsorption site (sulfur: fcc, phenyl: hcp), but also extends to most intricate structural details of the adsorption geometry. In the model depicted in Fig. 1 the optimized LEED and DFT structures cannot be distinguished by the naked eye. LEED derived vertical distances for each atom of the asymmetric unit of TMB are summarized in Fig. 1d. In addition to the deprotonation, TMB undergoes obvious structural changes upon adsorption. In the gas phase TMB is propeller-shaped due to steric hindrance between the σ-bonded phenyl rings, whereas in the adsorbed state this tilt is not present anymore. It is noteworthy, that for a structure simulation within the plane space group $p31m$, the chiral character of the propeller shape cannot be retained. However, a LEED structure refinement without any symmetry constraints likewise results in untwisted phenyl rings, in accordance with the DFT results.

In the following discussion of structural details quoted bond lengths and interatomic distances always refer to the LEED optimized structure, while DFT derived values are given in parentheses. Besides the removal of the propeller shape, a further prominent intra-molecular deformation in TMB is the short sulfur-copper distance, indicating covalent binding to the copper substrate. Sulfur is located above fcc hollow sites, but in a slightly asymmetric position closer to a twofold bridge site with similar distance to the two closer Cu atoms. The S-Cu distances amount to 2.64 [2.63] Å and 2.36 [2.37] Å, respectively. At least, the lower bond lengths are in good agreement with covalent S-Cu distances in CuS (2.19 – 2.38 Å)[69] and $Cu_2S$ (2.18 – 2.90 Å).[70] The copper atoms of the first layer adjacent to sulfur are lifted by 0.10 [0.07] Å with respect to the mean height of copper in the first layer (cf. ESI†, Table S4). These substrate relaxations can be seen as a consequence of covalent bond formation, as similarly found for tetracyanoquinodimethane on Cu(100).[71]

Also the organic backbone exhibits further slight deformations. The height of the carbon atom C6 (cf. Fig. 1 for numbering) is lower as a consequence of the downward bending of the sulfur atoms. This may result in a degradation of aromaticity, as induced by bond elongation and alternation,[72] as well as out-of-plane deformation,[73] with concomitant consequences for bonding properties in metal-organic networks. Both the central and peripheral phenyl rings are slightly distorted, as compared to the C-C bond length of 1.40 Å in benzene.[74] The outer phenyl rings exhibit deviating nearest neighbour C-C distances in the range of 1.39 –1.49 [1.41 – 1.43] Å. The hydrogen atoms in TMB are bent up with respect to the mean height of the carbon atoms, as also reported for benzene on Co(0001).[16] The next nearest neighbour C-C distance in the inner phenyl ring of TMB of 1.43 [1.42] Å is slightly elongated with respect to the gas phase [1.40 Å] (cf. ESI†, Table S2). This can be explained by stretching of TMB in order to simultaneously optimize all S-Cu bonds. This stretching is also noticeable in the C6-S distance. The value of the adsorbed molecule of 1.76 [1.79] Å is larger than in the gas phase [1.73 Å] (cf. ESI†, Table S2). Hence, stretching of the C6-S bond can be understood as a compromise between an optimal S-Cu bond length, without the necessity to reduce the distance between the aromatic system and the copper surface below its equilibrium value. The overall dimension of adsorbed TMB is expanded, as indicated by intramolecular S-S or C1-S distances of 13.09 [13.02] Å or 9.03 [8.98] Å, compared to 13.00 Å or 8.91 Å for the optimized trithiolate in the gas phase, respectively (cf. ESI†, Table S2).

Intermolecular S···H1 and S···H2 distances of 3.23 [3.32] Å and 2.85 [2.91] Å are comparatively large, thus intermolecular hydrogen bonds do not appear as an important contribution to the stabilization of the structure.[75]

It is also very instructive to compare optimized DFT structures obtained with and without dispersion correction. Including van der Waals interactions results in a significantly lower distance between the phenyl rings and the copper surface, *i.e.* the mean height of carbon decreases from 2.76 Å to 2.35 Å. This diminished adsorption height is in better agreement with the LEED result of 2.43 Å. Yet, neglecting the screening of van der Waals interactions through the free electrons of the metal support leads to overbinding as compared to the experimental results.

Nevertheless, the present results suggest that conventional dispersion corrected DFT yields more accurate results even on metal surfaces.

## Conclusions

We presented a combined experimental and theoretical structure refinement of the large trithiolate TMB on Cu(111). Out of six initially considered symmetry-allowed structures, the same model was clearly favoured by both LEED-I(V) and DFT. Both methods independently result in an adsorption geometry, where all sulfur atoms bind to fcc threefold hollow sites and all phenyl rings reside on hcp threefold hollow sites with every other carbon atom atop copper. This finally settles the question as to the preferred adsorption site of this molecule. In addition both techniques yield a wealth of further structural detail. The sulfur atoms are significantly moved down in order to establish a covalent bond with copper atoms. Sulfur does not adopt a fully symmetric position in the threefold hollow site, but remains closer to a twofold bridge site. The two adjacent copper atoms are also lifted from the substrate plane. Deformations of the organic backbone affect the planarity and the carbon-carbon distances in the phenyl rings. The remarkable agreement in these structural features obtained with the two independent techniques supports the conclusion that adsorption geometries of complex functional molecules can be accessed with sub-atomic precision. Besides the obvious power in the combination of the two techniques, the low experimental effort of LEED-I(V) experiments in comparison to synchrotron-based structural techniques is particularly appealing.

## Acknowledgements

This work was supported by the Nanosystems-Initiative Munich (NIM) funded by the Deutsche Forschungsgemeinschaft. TS acknowledges financial support by the Fonds der Chemischen Industrie (FCI).

## Notes and references

[a] Department of Physics, Technische Universität München, James-Franck-Str. 1, 85748 Garching, Germany and Center for NanoScience (CeNS), Schellingstr. 4, 80799 Munich, Germany
[b] Department of Chemistry, Technische Universität München, Lichtenbergstr. 4, 85747 Garching, Germany
[c] Center of Micro- and Nanochemistry and Engineering, Organische Chemie I, Universität Siegen, Adolf-Reichwein-Str. 2, 57068 Siegen, Germany
[d] Deutsches Museum, Museumsinsel 1, 80538 Munich, Germany
[e] Department of Earth and Environmental Sciences, Ludwig-Maximilians-University, Theresienstr. 41, 80333 Munich, Germany
[f] Department of Theoretical Physics, KTH Royal Institute of Technology, SE-106 91 Stockholm, Sweden
* E-mail: markus@lackinger.org; www.2d-materials.com

† Electronic Supplementary Information (ESI) available: Experimental and calculational details, constraints for LEED-I(V) optimization, complete I(V) dataset, optimization results of all competing structures and the additional adatom-based structures, selected atomic coordinates, xyz coordinates of optimized structures by LEED and DFT with and without dispersion correction. See DOI: 10.1039/b000000x/

**Table of contents entry**

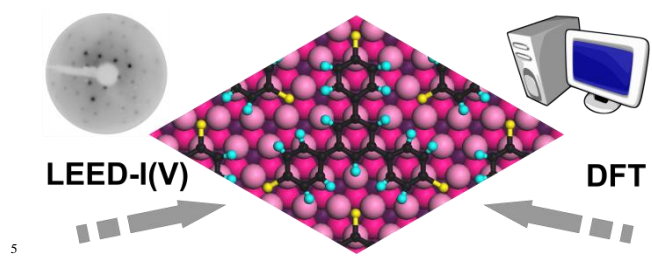

The surface geometry of Cu(111)/TMB is consistently yielded by LEED-I(V) and DFT-vdW. Structural details of molecular and upmost copper layers are analysed.